\def\opalabbiendi{OPAL Collaboration, G.\ Abbiendi \etal}

\newcommand{\PLB}[3]  {Phys.\ Lett.\ \textbf{B#1} (#2) #3}
\newcommand{\ZPC}[3]  {Z.\ Phys.\ \textbf{C#1} (#2) #3}
\newcommand{\EPC}[3]  {Eur.\ Phys.\ J.\ \textbf{C#1} (#2) #3}

\newcommand{\NPB}[3]  {Nucl.\ Phys.\ \textbf{B#1} (#2) #3}

\newcommand{\CPC}[3]  {Comp.\ Phys.\ Comm.\ \textbf{#1} (#2) #3}

\newcommand{\epem}{\mbox{$\mathrm{e^+ e^-}$}}

\newcommand{\WW} {\mbox{$\mathrm{W^+W^-}$}}

\newcommand{\WWg}{\mbox{\WW$\gamma$}}

\newcommand{\qq}{\mbox{$\mathrm{q\overline{q}}$}}

\newcommand{\lnu}{\mbox{$\ell\overline{\nu}_{\ell}$}}

\newcommand{\qqqq}{\mbox{\qq\qq}}

\newcommand{\qqlv}{\mbox{\qq\lnu}}
\newcommand{\qqln}{\mbox{\qq\lnu}}

\newcommand{\Mw}{\mbox{$M_{\mathrm{W}}$}}

\newcommand{\Gw}{\mbox{$\Gamma_{\mathrm{W}}$}}

\newcommand{\Egam}{\mbox{$E_\gamma$}}

\newcommand{\cosg}{\mbox{$\cos\theta_\gamma$}}

\newcommand{\GeV}{\mbox{$\mathrm{GeV}$}}

\newcommand{\roots}{\mbox{$\sqrt{s}$}}
\newcommand{\rootsprime}{\mbox{$\sqrt{s'}$}}
\newcommand{\sprime}{\mbox{${s'}$}}
\newcommand{\snorm}{\mbox{${s}$}}

\newcommand{\KoralW}{\mbox{KORALW}}
\newcommand{\KandY}{\mbox{KandY}}

\newcommand{\YFSWW}{\mbox{YFSWW}}

\def\etal{\mbox{{\it et al.}}}
\documentstyle[a4p,12pt,epsfig,amssymb]{article}
\begin{document}

\begin{titlepage}

\begin{center}
{\large }
\end{center}
\bigskip
\begin{flushright}
Cavendish-HEP-2003/08 \\ 
July 2003
\end{flushright}
\bigskip\bigskip
\begin{center}
 \Large{\bf \boldmath The Influence of the Experimental Methodology \\}
 \Large{\bf \boldmath on the QED Theoretical Uncertainties \\}
 \Large{\bf \boldmath on the Measurement of $\Mw$ at LEP}

\end{center}
\bigskip\bigskip\bigskip
\begin{center}
{ \large \bf M.\,A.\ Thomson \\}
\vspace{0.25cm}
{ \large \em Cavendish Laboratory, University of Cambridge, \\
 Cambridge CB3 0HE, England.}
\end{center}
\bigskip\bigskip\bigskip

\begin{center}{\large Abstract}\end{center}
Previous studies of the QED systematic uncertainties on the LEP measurement of 
the W-boson mass, $\Mw$, 
have used idealized event selections and fitting procedures. In this 
paper, the Monte Carlo tandem of KoralW and YFSWW is used to investigate how 
the full experimental $\Mw$ extraction procedure affects these estimates.
It is found that the kinematic fitting used in the experimental 
determination of $\Mw$ enhances the sensitivity to QED corrections involving 
real photon production. It is concluded that the previous estimates of the 
QED theoretical uncertainty on the LEP2 $\Mw$ measurement may be too small. 
A simple procedure for approximating the effect of the kinematic fit at the 
level of the generated four fermions is proposed. This procedure would
allow previous theoretical studies to be repeated using a much closer 
approximation of the real experimental mass extraction method. 
Finally,  the possibility of setting experimental limits on 
${\cal{O}}(\alpha)$ theoretical uncertainties using identified 
$\epem\rightarrow\WWg$ events is discussed.

\begin{center}
{\large\em Submitted to JHEP}
\end{center}
\end{titlepage} 

\section{Introduction}

It is anticipated that the final experimental error on the mass of the W boson,
$\Mw$, from LEP\,2 
will be approximately 35\,MeV. A potential source of theoretical uncertainty
is the treatment of QED and electroweak corrections in the Monte Carlo 
programs used to simulate the process $\epem\rightarrow4f(\gamma)$. 
In recent years ${\cal{O}}({\alpha})$ electroweak corrections have been 
implemented in the YFSWW\cite{bib:YFSWW} and 
RacoonWW\cite{bib:RacoonWW} programs. The current estimate of the theoretical 
systematic uncertainty on the LEP measurement of $\Mw$ due to these
predominantly QED corrections is less than 5\,MeV\cite{bib:sysmw}. However, 
as pointed out by the authors, this estimate is based 
on an idealized $\Mw$ extraction procedure. Indeed, the authors suggest
that the study should be repeated using the full experimental fitting
procedure. 

In this paper the concurrent Monte Carlo combination\cite{bib:KandY} 
of KoralW\cite{bib:KoralW} and YFSWW\cite{bib:YFSWW} 
(referred to as KandY) is used to study the sensitivity of the 
full experimental analysis to $\cal{O}(\alpha)$ corrections. 
Monte Carlo events generated using KandY are passed through the OPAL 
detector simulation\cite{bib:GOPAL} and W-boson reconstruction 
algorithm\cite{bib:MwO}. In this way, the previous estimates of 
the QED systematic uncertainties can be extended to the full 
experimental $\Mw$ extraction procedure. It is found that 
the sensitivity to the so-called\cite{bib:sysmw} $\cal{O}(\alpha)$
non-leading (NL) electroweak corrections is significantly enhanced by the 
kinematic fitting used in the experimental analyses. This enhancement is 
a result of the relatively large change in the cross section for production of
$\epem\rightarrow\WW\gamma$ events where the photon is away from the
region collinear with the $\epem$ beams. Consequently, measurements of the
cross section for $\epem\rightarrow\WW\gamma$ in this region place
constraints on the related $\Mw$ systematic uncertainties. 

It is suggested that previous 
studies\cite{bib:sysmw} be repeated using a closer approximation of the 
real experimental procedure. With this in mind, a simple procedure to 
approximate the effects of the kinematic fit at the level of the 
generated four fermions is proposed.

\section{\boldmath The Experimental Measurement of $\Mw$ at LEP\,2}

The main measurements of $\Mw$ at LEP are obtained 
from the direct reconstruction of the W-boson invariant mass
distribution in $\epem\rightarrow\qqln$ and $\epem\rightarrow\qqqq$ events.  
Although the exact methodologies used by the four LEP experiments 
differ\cite{bib:MwO,bib:MwADL}, there are many common 
features and the general principles of the analyses are summarized below. 

The experimental event selections for $\epem\rightarrow\WW$ have
high efficiency, typically $80\,\%-85\,\%$. Once events are 
selected, algorithms are applied to reconstruct the four-momenta of the 
fermions. 
In $\epem\rightarrow\qqln$ events, the charged particle tracks and any 
clusters in the calorimeters (electromagnetic and hadronic) which are not 
associated with the lepton are forced into two jets using the 
Durham\cite{bib:Durham} or LUCLUS\cite{bib:LUCLUS} algorithms.  
In $\epem\rightarrow\qqqq$ events, tracks and clusters are forced 
into four jets (although DELPHI and OPAL allow for the possibility
of a fifth `gluon' jet). The jet energy resolution is relatively poor, 
$\sigma_E/E > 60~\,\%/\sqrt{E}$, and neutrinos from leptonic W-boson 
decays are unobserved. Consequently,
kinematic fits are employed to improve the event-by-event mass resolution. The 
kinematic fits
impose the constraints of energy and momentum conservation. In most 
analyses the additional constraint of equal masses for the 
two W bosons is imposed, effectively neglecting $\Gw$, referred to as a 
five constraint
(5C) fit. The result of
the 5C kinematic fit is a single reconstructed mass for each event
which approximates to the average of the invariant masses of the 
fermion pairs from the two W-boson decays.
In general, photons from 
initial state radiation (ISR) within the detector acceptance 
are not treated as separate particles in the fit.
Although the four LEP collaborations employ different procedures
to extract $\Mw$ from the reconstructed mass distribution, each method
essentially corresponds to fitting the peak of the reconstructed
mass distribution and calibrating out experimental biases using 
Monte Carlo event samples.

The single most important aspect of the kinematic fit is the constraint that
the sum of the energies of the four fermions is equal to the centre-of-mass 
energy of the $\epem$ collision, $\roots$. Due to the poor jet energy
resolution, this constraint significantly improves the invariant mass 
resolution. 
However, in the presence of ISR, this procedure introduces a 
bias in the reconstructed mass, since the energies of the four fermions 
should be constrained to the centre-of-mass energy {\em after}
photon radiation, 
$\rootsprime$, rather than to $\roots$. As a result, for events with ISR, 
the kinematic fit will return a mass value which is too large. 
On average, the event-by-event mass returned by the kinematic fit 
approximately corresponds to    
\begin{eqnarray} \overline{M}& \approx &\frac{1}{2}(M_{12}+M_{34})\times\sqrt{{{s}\over{\sprime}}}\label{eqn:mass}.
\end{eqnarray}
Hence, the reconstructed invariant mass distribution depends
strongly on the $\rootsprime$ distribution and as a result 
the peak of reconstructed W-boson mass is approximately 100\,MeV 
higher than $\Mw$ (the exact value depending on $\roots$).
The bias in the fitted mass is even greater, typically a few hundred MeV.
This bias is subsequently determined using Monte Carlo events 
and the data measurement corrected. Consequently, deficiencies in the 
Monte Carlo simulation may result in systematic biases and it is
necessary for the Monte Carlo programs to predict accurately the 
$\rootsprime$ distribution.

\section{\boldmath QED Uncertainties on the LEP $\Mw$ Measurement}

The current estimate of 5\,MeV for the QED/Electroweak 
theoretical uncertainty on the 
LEP $\Mw$ measurement is based on the Monte Carlo generated
invariant mass distribution of the $\mu^-{\overline{\nu}}_\mu$ system in 
$\epem\rightarrow\mu^-{\overline{\nu}}_\mu u{\overline{d}}(\gamma)$ 
events\cite{bib:sysmw}. Uncertainties are obtained from the shifts 
in the invariant mass distribution which result from dropping higher order
theoretical terms. The simplest definition of invariant mass is obtained 
directly from the four-momenta of the generated fermions (termed 
{\em BARE} mass\footnote{The nomenclature of reference 
\cite{bib:sysmw} is used throughout this paper rather than that of
\cite{bib:LEP4F}.}). However, in the real experimental analyses,
photons from  final state radiation (FSR) 
are usually recombined with the 
nearest jet or lepton. This recombination is emulated at the 
level of the generated particles by recombining photons with the closest 
fermion if the photon has energy $\Egam<1$\,GeV or
if the invariant mass of the photon and fermion is less than  
5\,GeV(25\,GeV), termed the {\em CALO5}({\em CALO25}) photon
recombination scheme. 
In this way FSR photons are usually recombined with 
the appropriate fermion.  In the $CALO5/CALO25$ schemes photons within 
5$^\circ$ of the beam direction are treated as unobserved 
and are not recombined, reflecting the holes in experimental acceptance
along the beam direction. 

The main difference between the definitions of $BARE$ and $CALO$ masses 
and the experimentally reconstructed mass arises from the kinematic fit
employed in the experimental analyses. 
As a result of the kinematic fit, $\cal{O}(\alpha)$ QED corrections 
may affect the reconstructed mass in two ways: either by modifying 
the invariant mass distribution of the fermion pairs from W decays
or by modifying the $\rootsprime$ distribution. In the first case, 
the theoretical estimates of mass biases obtained using either 
{\em BARE} or {\em CALO} schemes provide a good estimate of the 
true experimental bias. However, in second case, even if the invariant 
mass distribution of the fermion pairs is unchanged, the experimentally 
measured mass distribution may be biased as a result of the kinematic fit 
scaling the measured energies to $\roots$ rather than $\rootsprime$.  
Consequently, previous estimates of the impact of theoretical corrections
which modify either the rate or energy spectrum of real photons produced 
in the process $\epem\rightarrow\WW\gamma$ may be too small. 

To study possible differences between the previous methods 
used to estimate $\Mw$ systematic uncertainties and the
true impact on the experimental measurement the Monte Carlo
tandem of \KoralW\ and \YFSWW\ (KandY) is used. A total of four million
$\epem\rightarrow\WW$
events was generated over 8 separate values of $\roots$ reflecting
the LEP\,2 data sample used in the W-mass determination ($183\,\GeV < \roots < 
209\,\GeV$).
The events were passed through the full OPAL detector 
simulation\cite{bib:GOPAL} and W-mass reconstruction procedure
including the 5C kinematic fit\cite{bib:mw172}. 
The following studies are based on events passing the
OPAL $\WW\rightarrow\qqln$ and  $\WW\rightarrow\qqqq$ event selections. 
The \KandY\ program produces correction weights enabling generated events to
be reweighted to the corresponding theoretical prediction  
removing certain terms. Correction weights were produced 
corresponding to: the replacement of the Screened Coulomb 
correction\cite{bib:Chapovsky} with the unscreened correction; degrading the 
treatment of ISR from ${\cal{O}}(\alpha^3)$ to  
${\cal{O}}(\alpha^2)$  YFS exponentiated\cite{bib:YFS} leading logarithm (LL); 
and dropping the so-called
${\cal{O}}(\alpha)$ NL electroweak corrections of KandY. 
It is important
to note that the corrections termed `${\cal{O}}(\alpha)$ NL' in 
\cite{bib:sysmw} include bremsstrahlung from the W bosons (referred to as
WSR) and the resulting interference with ISR. 

The treatment of the Coulomb correction modifies the invariant mass
distribution of the fermion pairs from the W decays. Consequently,
the shift in the $BARE/CALO$ invariant mass distribution provides
a good measure of the true experimental bias. Degrading the
treatment of ISR from ${\cal{O}}(\alpha^3)$ to  
${\cal{O}}(\alpha^2)$ exponentiated LL has negligible
impact on the experimental measurement of $\Mw$\cite{bib:MwO}
(the measurement is sensitive to ISR through a modification of
the $\rootsprime$ distribution, however the effects of the 
${\cal{O}}(\alpha^3)$ terms are small). 
The effect of the ${\cal{O}}(\alpha)$ NL electroweak 
corrections, which include bremsstrahlung from the W bosons, is more 
interesting. 
To study potential changes in distributions it is convenient to define the 
fractional change in a differential cross section,
\begin{eqnarray}
 \Delta_{NL} & = & {{(d\sigma_{KY}^{NL}-d \sigma_{KY})}\over{d\sigma_{KY}} },
\end{eqnarray}
where $d\sigma_{KY}$ is the differential cross section in a given
bin from the full $\KandY$ generation and $d\sigma_{KY}^{NL}$
is the corresponding distribution without the $\cal{O}(\alpha)$ NL
electroweak corrections.  

\begin{figure}[htbp]
\begin{center}
 \epsfxsize=18cm
 \epsffile{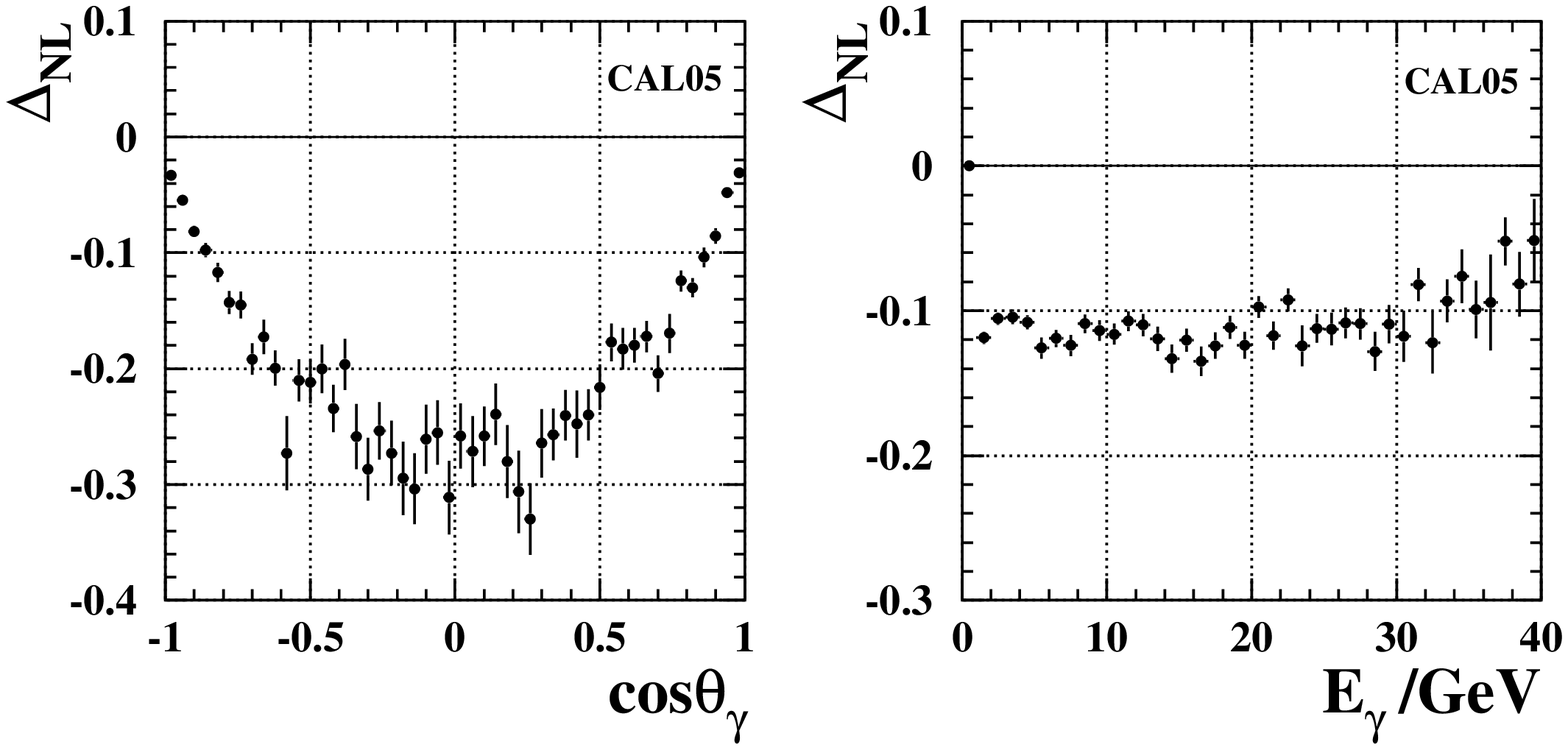}
 \caption{The effect of the ${\cal{O}}(\alpha)$ NL corrections in
          \KandY\ on the distributions of the energy and angle 
          with respect to the $e^-$ beam of the highest energy photon in
	  the event. Only photons which have not been recombined using 
          CALO5 scheme are included.       
 \label{fig:fig1}}
\end{center}
\end{figure}

Figure~\ref{fig:fig1} shows the effect of the $\cal{O}(\alpha)$ NL 
corrections on the photon energy and polar angle distribution. 
These plots are similar to those of Reference~\cite{bib:KandYFig9}
showing that the experimental selections do not introduce 
significant biases for events with additional photons\footnote{The plots
in Reference ~\cite{bib:KandYFig9} are for $\WW\rightarrow\qqlv\gamma$ 
events generated at $\roots=200\,\GeV$ and do not include the effects of
event selection.}.
The inclusion of the $\cal{O}(\alpha)$ NL corrections reduces the number 
of real photons produced in the detector acceptance.
The maximum fractional change occurs in the region perpendicular to
the beam axis where the $\WWg$
cross section is decreased by 30\,\%. This modification of the 
photon rate and angular distribution 
is predominantly due the inclusion of radiation from the W bosons. 
Specifically, the reduction in the $\WWg$ cross section
results from interference between ISR and radiation from the 
W bosons\footnote{This was verified using YFSWW. The reduction in 
the cross section for photon production shown in Figure~\ref{fig:fig1}a 
was reproduced by switching from {\tt KEYCOR=2} to {\tt KEYCOR=3} in YFSWW,
corresponding to switching from YFS exponentiation for ISR alone to 
the full YFS form factor including radiation from the W bosons (WSR) and 
interference between ISR and WSR.}. The photon energy spectrum is not greatly 
distorted by the NL corrections as can be seen from Figure~\ref{fig:fig1}. 

The potential bias in the measurement of $\Mw$ due to the
$\cal{O}(\alpha)$ NL terms is investigated using four definitions of 
reconstructed mass were considered:
\begin{itemize}
   \item ${\overline{M}}_{BARE}$ : the average of the two 
          W-boson masses determined from the four-momenta of
          the four fermions.
   \item ${\overline{M}}_{CALO5}$ : the average of the two 
          W-boson masses determined from the four-momenta of
          the four fermions after applying the $CALO5$ photon
          recombination procedure.
   \item ${M}_{5C}$ : the mass returned by the OPAL 5C kinematic
         fit. Only events for which the fit converges with a
         probability of greater than 0.1\,\% are plotted.
   \item ${\overline{M}}_{CALO5*}$ : ${\overline{M}}_{CALO5}$ rescaled
         by $\sqrt{{{\snorm}\over{\sprime}}}$ where $\rootsprime$ is the
         invariant mass of the four fermion system after recombination
         of photons: 
\end{itemize}
\begin{eqnarray} \overline{M}_{CALO5*}& = &\overline{M}_{CALO5}\times\sqrt{{{s}\over{\sprime}}}\label{eqn:calostar}.
\end{eqnarray}

\begin{figure}[htbp]
\begin{center}
 \epsfxsize=18cm
 \epsffile{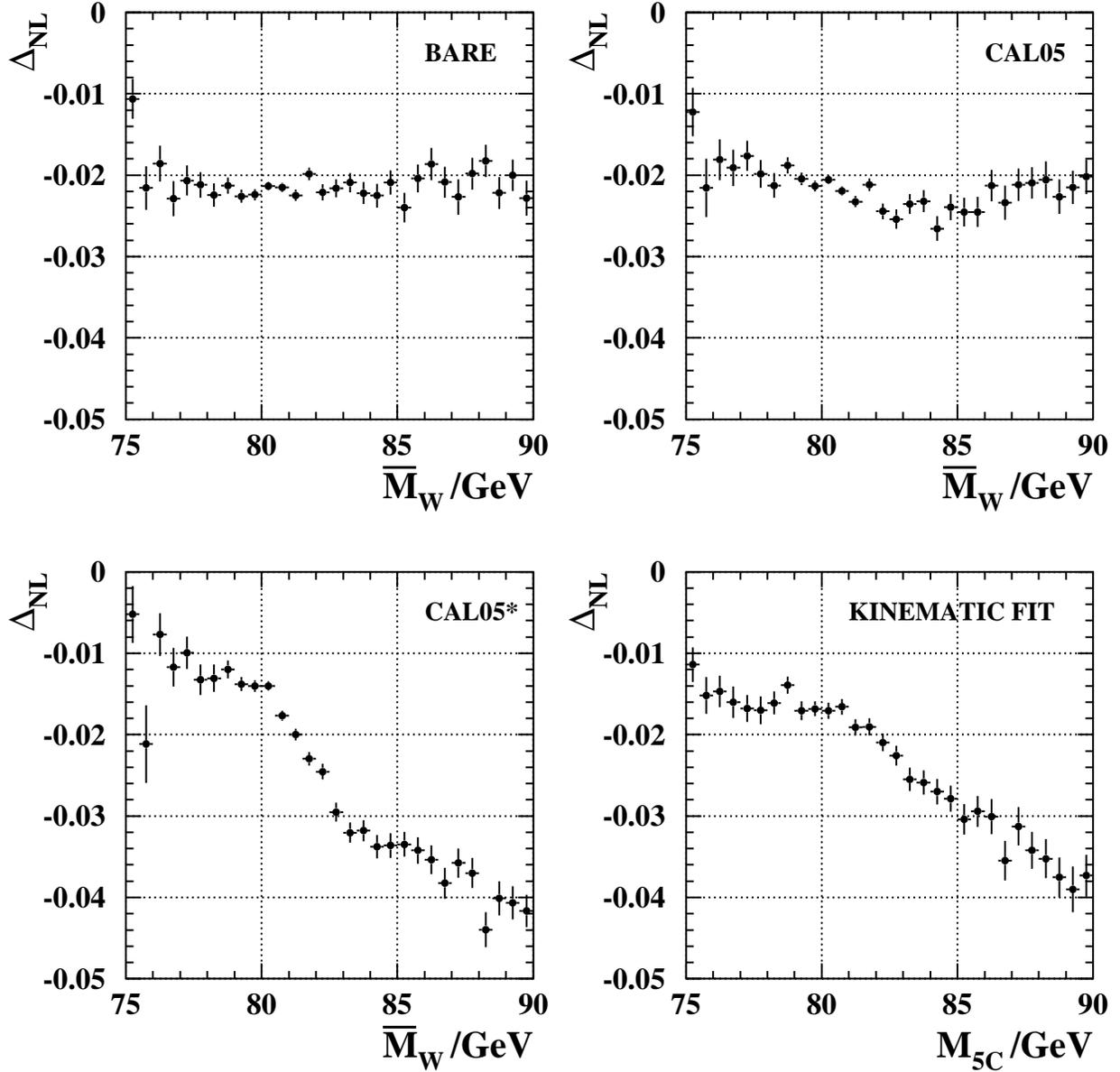}
 \caption{Fractional change in the differential cross-section as a function of
          the reconstructed mass using the $BARE$, $CALO5$, 
          and $CALO5*$ schemes compared to the experimental bias using
          the OPAL 5C kinematic fit. 
 \label{fig:fig2}}
\end{center}
\end{figure}

The $CALO5*$ definition, proposed in this paper, 
represents an attempt to emulate the effects
of the kinematic fit at the level of the four-momenta of the
fermions. The distortions of the W-mass peak, plotted for above definitions
of invariant mass, are shown in Figure~\ref{fig:fig2}. To quantify the
effect on the W-mass measurement, the resulting biases in the
mean reconstructed W-boson mass (in the region $75\,\GeV-90\,\GeV$) are listed in 
Table~\ref{tab:bias}. 
The inclusion of the  $\cal{O}(\alpha)$ NL corrections has almost 
no effect on the
$BARE$ mass distribution and the mean value changes by just
1\,MeV. A small but visible effect is observed for the mass in the 
$CALO5$ scheme giving a bias in the mean mass of 3\,MeV. These results
are consistent with those presented in \cite{bib:sysmw}.
However, the inclusion of the $\cal{O}(\alpha)$ NL corrections 
produces a significant shift in the experimentally reconstructed mass 
from the 5C kinematic fit, with the mean reconstructed mass in the 
range $75\,\GeV-90\,\GeV$ shifted by 21\,MeV. Although the invariant mass
distribution of the fermion pairs is almost unaffected the
reconstructed mass distribution is significantly biased. 
The origin of this modification of the $\rootsprime$ distribution
is the reduction in the production cross section for $\epem\rightarrow\WWg$,
which results from the interference of ISR and radiation from the W bosons.
It should be noted that by scaling the mean mass in the CALO5 scheme by
$\sqrt{s/s^\prime}$ (denoted $CALO5*$) the effect of the kinematic fit 
can be approximated at the level of the generated fermions. 

\begin{table}[htbp]
\renewcommand{\arraystretch}{1.1}
\begin{center}
\begin{tabular}{|l|r|} \hline 
     Scheme              & $\Delta M^{NL}$   \\ \hline
   {\em BARE}            &   +1\,MeV          \\
   {\em CALO5}           &   +3\,MeV          \\
   {$ CALO5*$}           &   +22\,MeV         \\
   {\em 5C FIT}          &   +21\,MeV         \\ \hline
\end{tabular}
\end{center}
\caption{The change in the mean W-boson mass in the range $75\,\GeV-90\,\GeV$ 
         when dropping the electroweak $\cal{O}(\alpha)$ NL 
         corrections from \KandY.
         Shifts are shown for three different theoretical definitions
         of the event-by-event reconstructed mass: {\em BARE}, {\em CALO5}
         and the new definition proposed in this paper, {$CALO5*$}.
         {\em 5C FIT} refers to the shift in the average of reconstructed
         mass distribution for the full OPAL event reconstruction using
         a 5C kinematic fit. 
\label{tab:bias} }
\end{table}

The effect of the kinematic fit is to make the experimental analyses
more sensitive to theoretical corrections which modify the
rate/spectrum of photons produced in the process $\epem\rightarrow\WWg$.
The bias of 21\,MeV associated with the ${\cal{O}}(\alpha)$ NL
corrections in KandY is almost an order of magnitude larger than previously quoted.
Since the origin of the 21\,MeV bias is interference between ISR and radiation
from the W bosons, one might expect this to be  
well accounted for in YFSWW, but this requires further evaluation. 
It should be noted that the 21\,MeV bias would be reduced significantly
if observed photons were treated as additional particles in the kinematic
fits. In this case the energies of the fermions from W decay would be
scaled correctly to $\rootsprime$.
  
For the OPAL W-mass extraction procedure the effect of 
the ${\cal{O}}(\alpha)$ NL corrections
on the full $\Mw$ reconstruction procedure is somewhat smaller than the
shift in the average event-by-event reconstructed mass quoted above.
Using a modified version of the OPAL $\Mw$ convolution fit\cite{bib:MwO}
the bias associated with dropping the ${\cal{O}}(\alpha)$ NL corrections 
is 14\,MeV.

\section{\boldmath Limits from the Measurement of the $\epem\rightarrow\WWg$ 
        Cross Section }

Measurements of the process $\epem\rightarrow\WWg$ have been performed at 
LEP\cite{bib:OPALwwg,bib:L3wwg,bib:DELPHIwwg}. 
These measurements, which are restricted to photons 
within the experimental acceptance (typically $|\cosg|<0.95$),
provide a direct probe of ${\cal{O}}(\alpha)$ QED corrections which 
affect real photon production away from the region collinear with the
$\epem$ beams. Of particular interest here are the  ${\cal{O}}(\alpha)$ NL 
electroweak corrections of KandY which decrease the calculated
$\epem\rightarrow\WW\gamma$ cross section. 
The largest change in the differential $\WWg$ cross section occurs 
close to the beam direction (as would be expected from ISR-WSR 
interference). However, when integrated over the photon polar angle, 
88\,\% of the change in cross section occurs within the nominal 
experimental acceptance of $|\cosg|<0.95$. 
Hence,  by measuring the $\epem\rightarrow\WW\gamma$ cross section for 
events where the photon is observed, it is possible to probe experimentally 
the ${\cal{O}}(\alpha)$ NL corrections of KandY. 

When the four LEP experiments publish final results
including all LEP\,2 data, the combined experimental precision on the 
$\WWg$ cross section should be approximately 7\,\%. 
The inclusion of the ${\cal{O}}(\alpha)$ NL electroweak corrections 
of KandY produces a $20\,\%$ decrease in the 
$\epem\rightarrow\WW\gamma$  cross section for $|\cosg|<0.95$ and
$\Egam>5$\,GeV (cuts chosen to match the experimental acceptances
of the LEP detectors). Thus, using rate alone, 
the measurements are of sufficient statistical 
precision to provide a useful test of the KandY 
implementation of WSR and in particular of WSR$-$ISR interference. 
Agreement with the predictions of KandY would provide the first 
indirect observation (albeit only at the 3 $\sigma$ level) of real 
photon radiation from the W bosons through interference with ISR diagrams. 
The statistical precision of this test would be improved slightly by 
including angular information in addition to the photon rate. 

In the context of the $\Mw$ measurement at LEP the 
inclusion of the  ${\cal{O}}(\alpha)$ NL corrections in KandY 
produces a 21\,MeV shift in the mean reconstructed mass which results
from the modification of the $\rootsprime$ distribution. This is a 
direct consequence of the decrease in the $\epem\rightarrow\WW\gamma$ 
cross section. Therefore, measurements of
the  $\epem\rightarrow\WW\gamma$ cross section at LEP provide constraints
on the ${\cal{O}}(\alpha)$ QED uncertainties related to real 
photon production away from the region  collinear with the beams. 
A 20\,\% reduction in the $\WWg$ cross section corresponds to a 21\,MeV 
shift in the reconstructed value of $\Mw$. 
Assuming the bias scales linearly, a measurement of 
the $\WWg$ cross section with a statistical precision of $7$\,\% 
would correspond to approximately 7\,MeV on the associated 
systematic uncertainty on $\Mw$. It should be noted that this limit would
apply to all theoretical corrections to the process $\epem\rightarrow\WW$
which modify the rate of real photon production away from the 
collinear region; it is not specific to the  
${\cal{O}}(\alpha)$ NL corrections of KandY.
Given that the expected  statistical uncertainty on the LEP 
$\Mw$ measurement is approximately 35\,MeV, a LEP combined measurement of
the $\WWg$ cross section would provide a useful experimental constraint
on the ${\cal{O}}(\alpha)$ QED systematic uncertainty.

\section{Conclusions}

Previous studies of QED systematic uncertainties on the LEP measurement of 
$\Mw$ have used idealized event selections and fitting procedures. 
In this paper the KandY Monte Carlo program was used to study the
effect of the experimental methodology on the estimated QED systematic
uncertainty. It is found that the kinematic fitting used in the experimental 
determination of $\Mw$ enhances the sensitivity to all QED corrections 
involving real photon production and it is concluded that 
previous estimates of the QED theoretical uncertainty on the LEP\,2 $\Mw$ 
measurement may be too small.  
A simple procedure for approximating the effect of the kinematic fit at the 
level of the generated four fermions is proposed. This procedure would
allow previous theoretical studies to be extended to include 
a good approximation of the real experimental mass extraction method. 
Finally,  it is shown that useful experimental limits on  
${\cal{O}}(\alpha)$ theoretical uncertainties related to real photon
production away from region collinear with the $\epem$ beams may
be inferred from LEP measurements of the 
$\epem\rightarrow\WWg$ cross section.

\section{Acknowledgments}

I would like acknowledge the authors of the \KandY\ program, S.~Jadach, 
W.~P{\l}aczek, M.~Skrzypek, B.F.L.~Ward and Z.~W\c{a}s, without whom this 
study would not have been possible. In particular I thank 
M.~Skrzypek and W.~P{\l}aczek for their many useful comments and suggestions.
I also would like to thank my colleagues on the 
OPAL collaboration for allowing me to present results using the 
OPAL detector simulation and W mass analysis tools. In particular 
M.~Verzocchi for his work in generating the \KandY\ Monte Carlo 
samples and R.~Hawkings for providing results from the full OPAL $\Mw$
fit. Finally I thank D.R.~Ward and C.P.~Ward for reading through draft
versions of this paper.

\newpage

\end{document}